\documentclass[prd,showpacs,nofootinbib,amsmath,amssymb,floatfix,superscriptaddress,showkeys]{revtex4}
\usepackage{multirow}
\usepackage{graphicx}% Include figure files
\usepackage{epstopdf}
\usepackage{dcolumn}% Align table columns on decimal point
\usepackage{bm}% bold math
\usepackage{threeparttable}
\usepackage{subfigure}
\usepackage{color,txfonts}
\usepackage{ulem}
\definecolor{blue0}{rgb}{0,0,0.6}
\usepackage[colorlinks,linkcolor=blue0,anchorcolor=blue0,citecolor=blue0,urlcolor=blue0]{hyperref}

\newcommand{\jcap}{J. Cosmol. Astropart. Phys.}

\newcommand{\mnras}{Mon. Not. R. Astron. Soc.}

\newcommand{\beq}{\begin{equation}}
\newcommand{\eeq}{\end{equation}}
\newcommand{\beqa}{\begin{eqnarray}}
\newcommand{\eeqa}{\end{eqnarray}}

\begin{document}

\title{Using $\gamma$-ray observations of dwarf spheroidal galaxies to test the possible common origin of the W-boson mass anomaly and the GeV $\gamma$-ray/antiproton excesses}

\author{Ben-Yang Zhu}
\affiliation{Guangxi Key Laboratory for Relativistic Astrophysics, School of Physical Science and Technology, Guangxi University, Nanning 530004, China}
\author{Shang Li}
\affiliation{School of Physics and Optoelectronics Engineering, Anhui University, Hefei 230601, China}
\author{Ji-Gui Cheng}
\affiliation{Guangxi Key Laboratory for Relativistic Astrophysics, School of Physical Science and Technology, Guangxi University, Nanning 530004, China}
\author{Xiao-Song Hu}
\affiliation{Guangxi Key Laboratory for Relativistic Astrophysics, School of Physical Science and Technology, Guangxi University, Nanning 530004, China}
\author{Rong-Lan Li}
\affiliation{Guangxi Key Laboratory for Relativistic Astrophysics, School of Physical Science and Technology, Guangxi University, Nanning 530004, China}
\author{Yun-Feng Liang}
\email[]{liangyf@gxu.edu.cn}
\affiliation{Guangxi Key Laboratory for Relativistic Astrophysics, School of Physical Science and Technology, Guangxi University, Nanning 530004, China}

\date{\today}

\begin{abstract}
A recent result from Fermilab suggests that the measured W-boson mass deviates from the prediction of the Standard Model (SM) with a significance of $>7\sigma$, and there may exist new physics beyond the SM. It is proposed that the inert two Higgs doublet model (i2HDM) can well explain the new W-boson mass. Meanwhile, the lightest neutral scalar $S$ in the i2HDM can be stable and play the role of dark matter with a preferred dark matter mass of $\sim 54-74$ GeV. It is also found that part of the parameter space of this model can explain both the Galactic center GeV gamma-ray excess detected by $Fermi$-LAT and the GeV antiproton excess detected by AMS-02 through a $SS\rightarrow WW^*$ annihilation. In this paper, we aim to test the possible common i2HDM origin of the three anomaly/excesses using the $Fermi$-LAT observations of Milky Way dwarf spheroidal (dSph) galaxies.
We perform single and stacking analyses on 19 dSphs that have J-factor measurements.
We find that our upper limits are below the favored parameters and seems to be able to exclude the possibility of a common origin of the three anomaly/excesses. However, because the J-factor measurements include relatively large uncertainties, which come from the measurements of stellar kinematics, whether this model could be reliably excluded needs to be further confirmed by future observations.
\end{abstract}
%\pacs{95.35.+d, 95.85.Pw, 98.52.Wz}

\maketitle

\section{Introduction}

The presence of dark matter (DM) particles in the universe is supported by many astrophysical observations. 
Latest observation results show that non-baryonic cold DM contributes $\sim$ 84\% of the matter density of the Universe \cite{Ade:2015xua}. 
However, the particle nature of the DM is still unknown. 
The most promising DM candidates are weakly interacting massive particles (WIMPs). 
WIMPs can produce $\gamma$-ray or cosmic-ray signals through annihilating or decaying into the Standard Model (SM) particles \cite{Jungman:1995df, Bertone:2004pz, Hooper:2007qk, Feng:2010gw,PhysRevD.69.123501}, which provide a possible approach for probing DM particles.
With the astrophysical instruments such as the $Fermi$ Large Area Telescope ($Fermi$-LAT \cite{atwood09lat,charles16review}) and the Dark Matter Particle Explorer (DAMPE \cite{DAMPE:2017,Chang:2017n,ltc22dampe}), a lot of efforts have been made to search for WIMP DM.

The Collider Detector at Fermilab (CDF) collaboration recently reported an exciting progress in physics, providing an informative hint for the study of DM.
They reported their newly measured W-boson mass $m_{\rm W}=80.4335\pm0.0094\,{\rm GeV}$ \cite{wboson}, which deviates from the prediction of the SM by $>7\sigma$ (note however that the measurements by
Refs.~\cite{ALEPH:2013dgf,ATLAS:2017rzl,LHCb:2021bjt} slightly conflict with this new measurement). Such a large discrepancy provides strong evidence of the presence of new physics. 
The most straightforward explanation for the mass anomaly is the existence of new particles/fields interacting with W-boson, which can enhance the W-boson mass through a radiative correction.
Various new particles and/or fields have been introduced to interpret the W-boson mass anomaly \cite{Athron22,lu22,ttp22,ygw22,Du:2022pbp,Yang:2022gvz,Strumia:2022qkt,Endo:2022kiw,Han:2022juu,Ahn:2022xeq,Zheng:2022irz,Perez:2022uil,Zhang:2022nnh,Borah:2022obi,Zeng:2022lkk,Du:2022fqv}. 
As one of the simplest models, Ref. \cite{fyz22} proposed that the inert two Higgs doublet model (i2HDM) can interpret the new W-boson mass without violating other astrophysical/experimental constraints.
% \cite{}.
More intriguingly, in this model the lightest new scalar $S$ is stable and can be a dark matter particle (more details of the model are described in Sec.~\ref{sec:model}), the annihilation of which through $SS\rightarrow b\bar{b}$ and $SS\rightarrow WW^*$ channels will produce antiprotons and gamma rays \cite{fyz22,zcr22}, and can simultaneously explain the Galactic center (GC) GeV excess \cite{Hooper:2010mq,Gordon:2013vta,Hooper:2013rwa,Daylan:2014rsa,Zhou:2014lva,Calore:2014xka,fermi17GCE} and antiproton excess signals \cite{cui17antiproton,cuoco17antiproton}.
The consistency of the DM particle parameters required to account for the three anomaly signals suggests a possible common origin of them.

The kinematic observations show that dwarf spheroidal galaxies (dSphs) are DM-dominated systems. Besides the GC, the dSphs are another most promising targets for indirect detection of DM due to their vicinity and low gamma-ray background \cite{Lake:1990du,Baltz:2004bb,Strigari:2013iaa}. Previously, based on the non-detection of a robust gamma-ray signal in the direction of dSphs, people have set very strong constraints on the mass $m_\chi$ and the annihilation cross section $\left<\sigma v\right>$ of the particle DM \cite{fermi11dsph,GeringerSameth:2011iw,2012PhRvD86b3528C,tsai13dsph,fermi14dsph,zhao2016ds,gs15dsph,fermi15dsph,hoof20}. In this work, we aim to test the i2HDM model with the $Fermi$-LAT observations of the Galactic dSphs. We only consider the dSphs that have J-factor measurements. 
The sample used for analysis are listed in Table \ref{tab1}.
We search for gamma-ray signals from these sources and then compare the results with the model-predicted gamma-ray flux from the $SS\rightarrow WW^*$ annihilation. 
To enhance the sensitivity, we also apply a stacking method (not the same as the commonly used combined likelihood analysis \cite{fermi11dsph,tsai13dsph,fermi14dsph,fermi15dsph}) to simultaneously consider all our sources together.
We do not find any signals from these dSphs and give constraints on  the parameter of $\left<\sigma v\right>_{SS\rightarrow WW^*}$.

\section{Inert two Higgs doublet model}
\label{sec:model}
The inert two Higgs doublet model is a minimal SM extension that introduces an additional Higgs doublet $H_2$ on the basis of the Standard Model \cite{Deshpande:1977rw,Ma:2006km,Barbieri:2006dq,LopezHonorez:2006gr}. In contrast, there exists only one Higgs doublet $H_1$ in the SM, i.e.
\begin{equation}
H_1=\left[\begin{array}{c}
G^{+} \\
\frac{1}{\sqrt{2}}\left(v+h+i G^0\right)
\end{array}\right], \quad H_2=\left[\begin{array}{c}
H^{+} \\
\frac{1}{\sqrt{2}}(S+i A)
\end{array}\right] .
\end{equation}
where $G^\pm$, $G^0$ are the charged and neutral Goldstone bosons, $h$ is the SM Higgs field and $v \approx 246\,{\rm GeV}$ is its vacuum expectation value. In addition to the SM Higgs, four new particles beyond the SM are introduced, namely the two neutral Higgs bosons $S$ and $A$, and a pair of charged Higgs $H^\pm$. The most general $\rm SU(2)\times U(1)$ gauge-invariant and renormalizable scalar potential of i2HDM can be written as
\begin{equation}
\begin{aligned}
V= & \mu_1^2\left|H_1\right|^2+\mu_2^2\left|H_2\right|^2+\lambda_1\left|H_1\right|^4+\lambda_2\left|H_2\right|^4 +\lambda_3\left|H_1\right|^2\left|H_2\right|^2+ \\
& \lambda_4\left|H_1^{\dagger} H_2\right|^2 +\frac{\lambda_5}{2}\left[ (H_1^{\dagger} H_2)^2+(H_1 H_2^{\dagger})^2\right]
\end{aligned}
\end{equation}
with $\mu_2$, $\lambda_2$, $\lambda_3$, $\lambda_4$, $\lambda_5$ the 5 free parameters.

The masses of electroweak gauge bosons in the SM are produced by the spontaneous symmetry breaking. When the additional scalar doublet $H_2$ is introduced, the non-SM scalar can increase the W-boson mass by a loop correction, and thus can explain the W-boson mass anomaly of CDF II (the mass anomaly can also be explained by introducing other new particles beyond the SM, e.g., axion-like particles, dark photons \cite{ygw22}). In the i2HDM model, the W-boson mass $m_W$ is related to the one in the SM $m_{W,{\rm SM}}$ by \cite{Peskin:1991sw,Eriksson:2009ws}
\begin{equation}
m_W^2=m_{W, \mathrm{SM}}^2+\frac{\alpha c_W^2 m_Z^2}{c_W^2-s_W^2}\left[-\frac{S}{2}+c_W^2 T+\frac{c_W^2+s_W^2}{4 s_W^2} U\right]
\end{equation}
where $c_W=\cos{\theta_W}$ and $s_W=\sin{\theta_W}$ with $\theta_W$ the Weinberg angle, $\alpha$ is the fine structure constant and $m_Z$ the Z boson mass. The oblique parameters $\mathcal{S}$, $\mathcal{T}$, $\mathcal{U}$ parameterize the contributions of possible new physics beyond the SM to electroweak radiative corrections. 
The i2HDM model has been shown to be able to explain both the CDF II W-boson mass without violating various existing astrophysical and experimental constraints \cite{fyz22}. The allowed model parameters on the $\left<\sigma v\right>-m_S$ panel are shown in Fig.~\ref{fig2} of Sec.~\ref{sec:constr}.

Another motivation for the i2HDM model to introduce an additional Higgs doublet is to interpret the dark matter. The two neutral scalars $S$ and $A$ contained in the second inert doublet can serve as the role of DM. Consider the existence of a discrete $Z_2$ symmetry, which has two eigenstates: $Z_2$-even (eigenvalue $+1$) and $Z_2$-odd (eigenvalue $-1$), assuming that all particles in the SM are $Z_2$-even ($+1$), and that the lightest new particle beyond the SM is $Z_2$-odd ($-1$), then the new particle is not able to decay into a $Z_2$-even SM particle and is therefore stable, providing a natural candidate for dark matter. 
In agreement with Ref.~\cite{fyz22}, the scalar $S$ is assumed to be lighter and therefore a DM particle (hence in the latter part of this paper the symbols $\chi$ and $S$ denote the same particle). Due to the symmetry of $S$ and $A$, the results will not change if $A$ is the DM \cite{fyz22,Arhrib:2013ela,Tsai:2019eqi}. The $S$ particles in the DM halo of the Milky Way can annihilate producing gamma rays or cosmic-ray particles. Annihilation through the $SS\rightarrow WW^*$ or $SS\rightarrow b\bar{b}$ channels is found to be able to explain the GeV gamma-ray excess in the Galactic Center as well as the possible GeV antiproton excess \cite{zcr22}. More interestingly, the allowed parameters overlap with some of the parameters that can explain the W-boson mass anomaly, which suggests that they may have a common origin. The purpose of this paper is to use the {\it Fermi}-LAT observations of dSphs to test whether these overlapping parameters can be supported by the dSph observations.

	\begin{figure*}
		\includegraphics[width=0.39\textwidth]{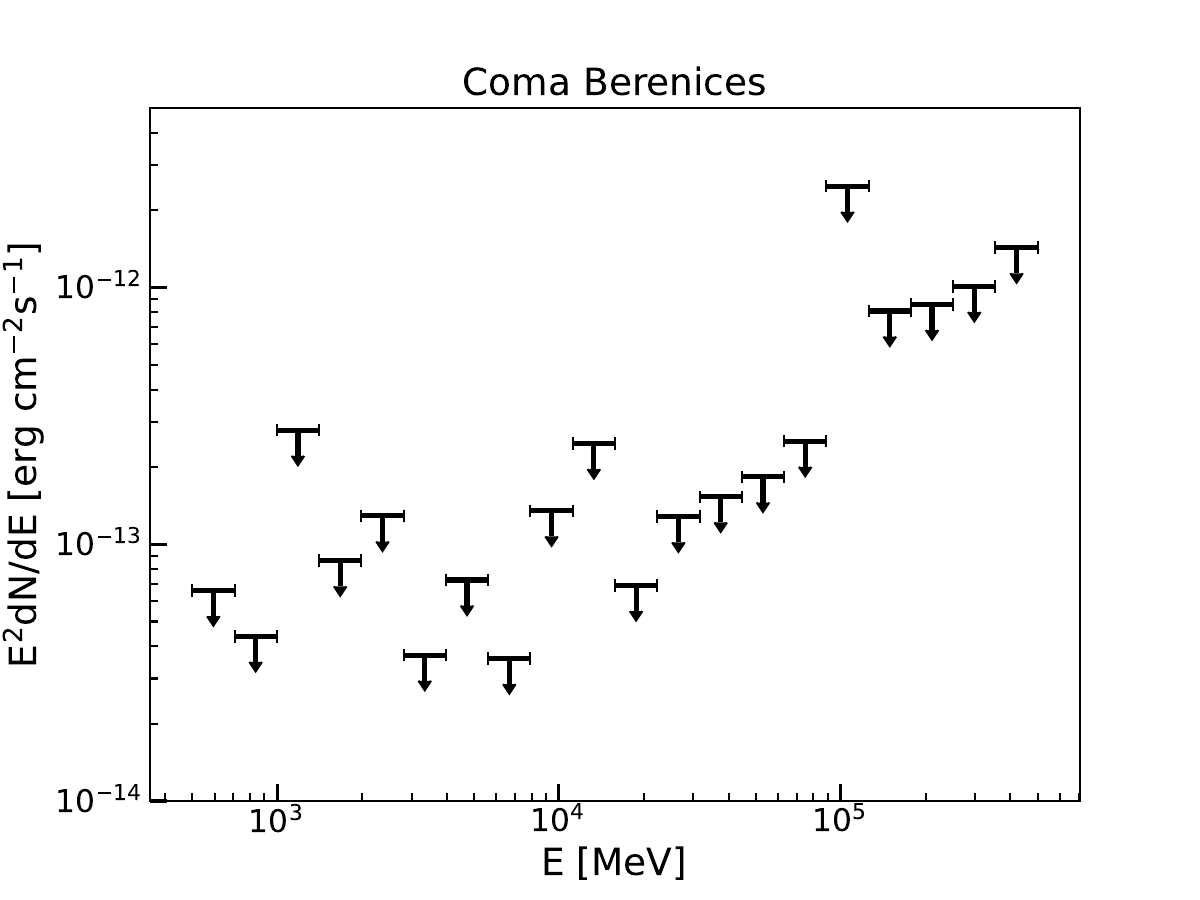}
		\includegraphics[width=0.39\textwidth]{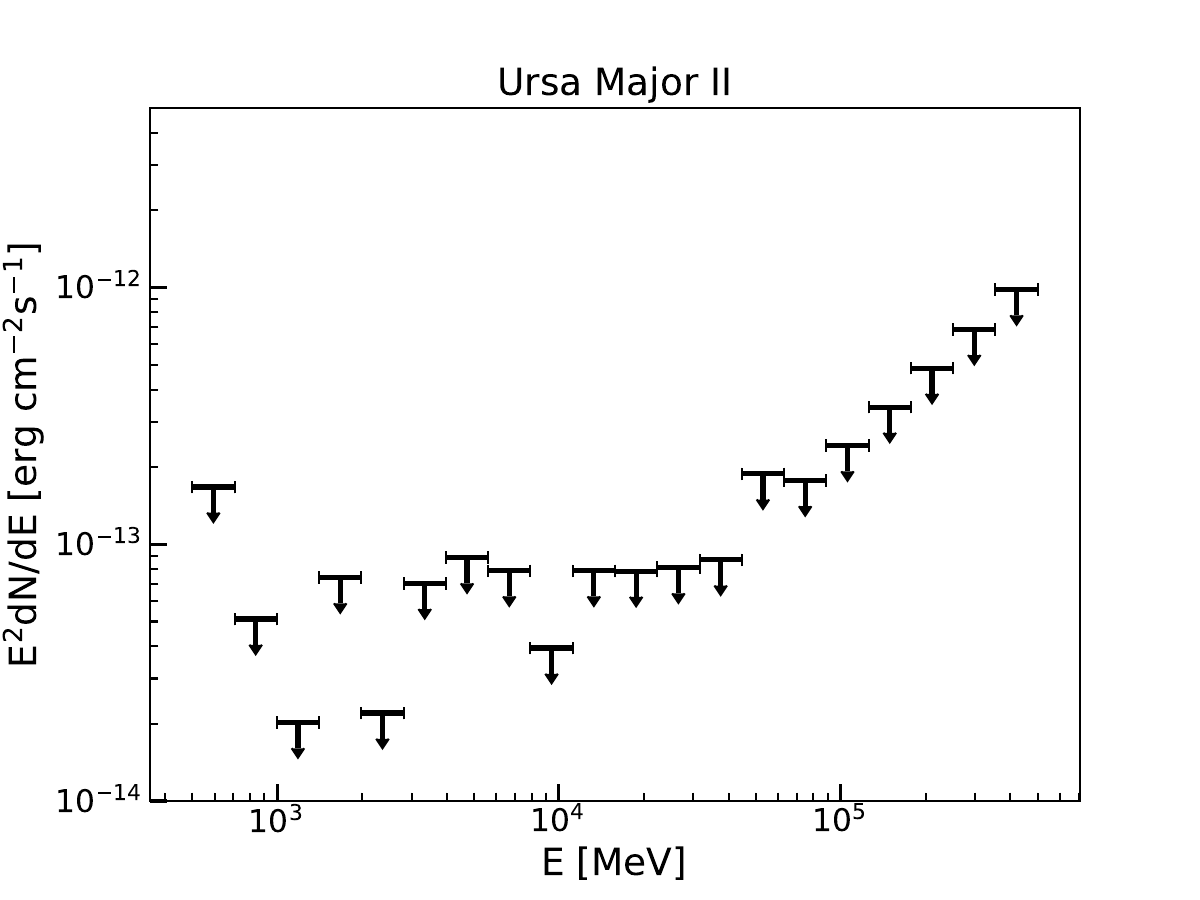}\\
		\includegraphics[width=0.39\textwidth]{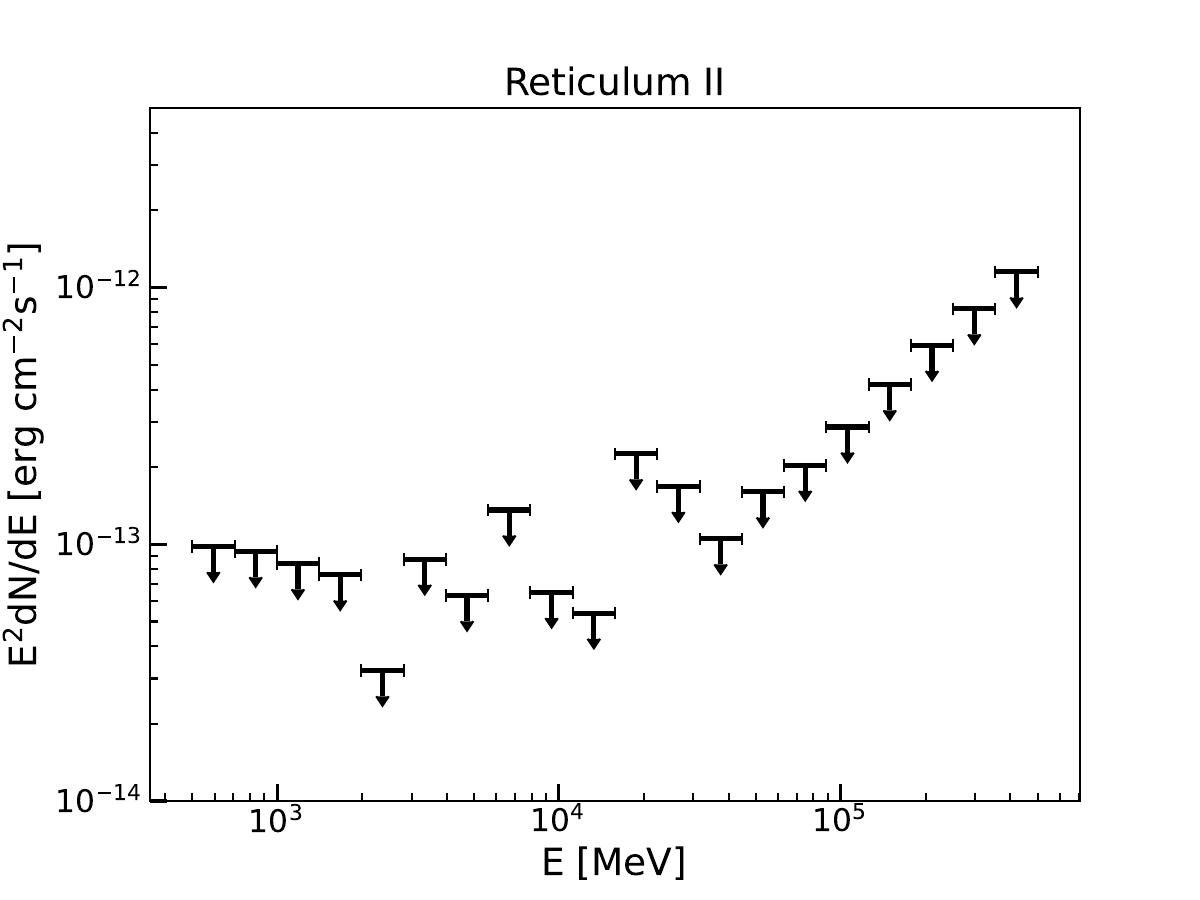}
		\includegraphics[width=0.39\textwidth]{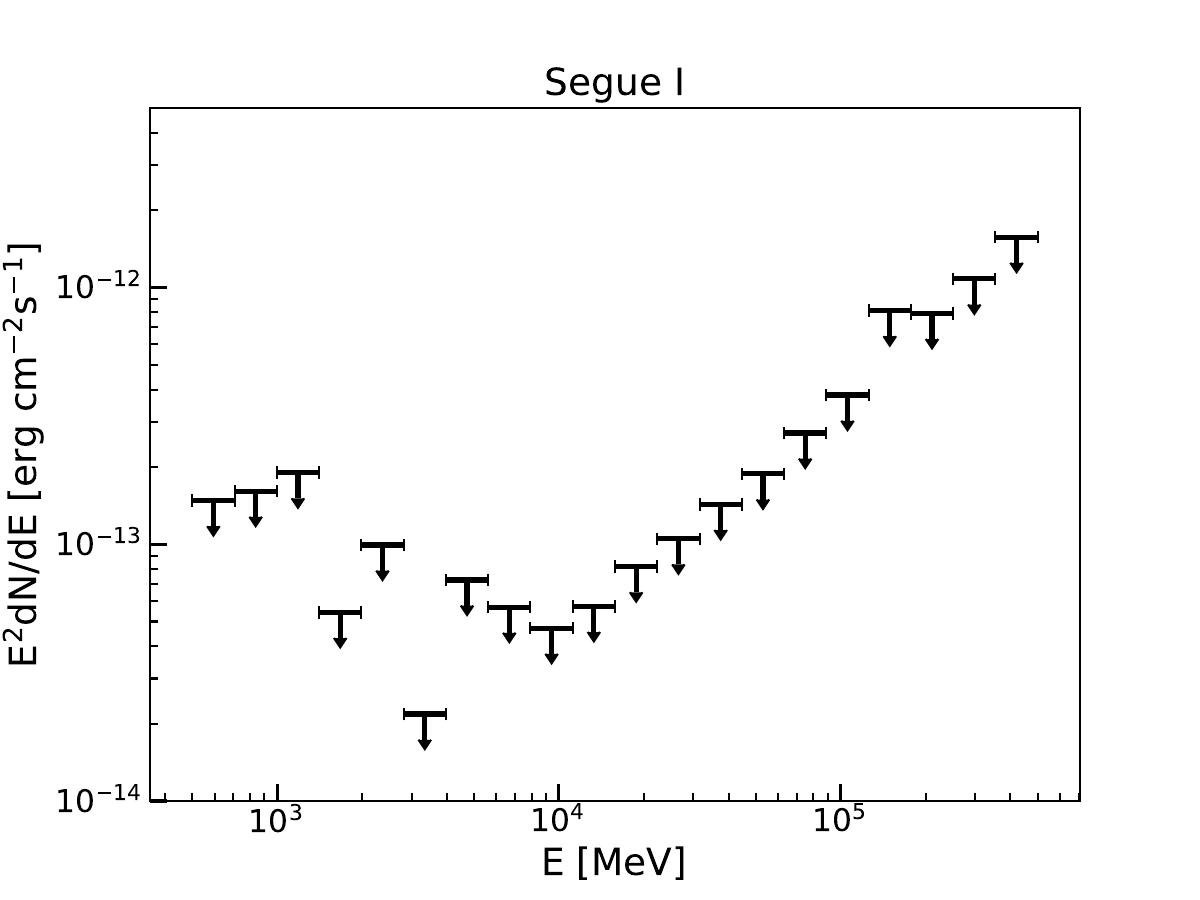 }
        \caption{Bin-by-bin flux upper limits at a 95\% confidence level for the 4 dwarf spheroidal galaxies with the biggest J-factors considered in this work. For other sources, please see Fig.~\ref{figA2}.}
		\label{fig1}
	\end{figure*}

\section{Searching for dark matter emission from the dSphs}
\label{sec:analysis}
We use $>$14 years (i.e. MET 239557417 - 705811348, from 2008 October 27 to 2023 May 15) of $Fermi$-LAT Pass 8 data in the energy range of 500 MeV to 500 GeV. To remove the Earth's limb emission, we use only the $\gamma$ events with zenith angle $<100^\circ$. Meanwhile, the quality-filter cuts ({\tt DATA\_QUAL==1 \&\& LAT\_CONFIG==1}) are applied to ensure the data valid for scientific analysis. We take a $15^{\circ}\times 15^\circ$ region of interest (ROI) for each target to perform a binned analysis. The latest version (Ver. 2.2.0) of {\tt Fermitools} is used to analyze the $Fermi$-LAT data. To model the background, we consider all 4FGL-DR3 \footnote{\url{https://fermi.gsfc.nasa.gov/ssc/data/access/lat/10yr_catalog/}} 
sources \cite{4fglA} within a {15$^{\circ}$} circle region centered on each target and two diffuse models (the Galactic diffuse gamma-ray emission {\tt gll\_iem\_v07.fits} and the isotropic component {\tt iso\_P8R3\_SOURCE\_V3\_v1.txt}).
All  dSphs are modeled as point-like sources.

We first perform a standard binned likelihood analysis \footnote{\url{https://fermi.gsfc.nasa.gov/ssc/data/analysis/scitools/binned_likelihood_tutorial.html}} to obtain the best-fit background model. During the fit, the parameters of all 4FGL-DR3 sources within the ROI, together with the re-scale factor of the two diffuse components are set free. On the basis of the best-fit background, we search for gamma-ray emission from the dSphs and derive the test-statistic (TS) values and flux upper limits of the targets. 
%The analysis method here is similar to that developed in \cite{fermi11dsph,tsai13dsph,fermi14dsph} and more details can be found in these articles. 
The TS is a quantity used to quantify the significance of the putative gamma-ray emission from dSph and is defined as ${\rm TS}=2\ln({L_{\rm dsph}}/{L_{\rm bkg}})$ \cite{1996ApJ...461..396M}, where the $L_{\rm bkg}$ and $L_{\rm dsph}$ are the best-fit likelihood values for the background-only model and the model containing a putative dSph, respectively. The significance is roughly the $\sqrt{\rm TS}$.
%Assuming a power-law spectrum with variable spectral index for the dSph emission, the TS values obtained in our analysis are listed in Table~\ref{tab1}.
%As is shown, no significant signal ($\rm TS>25$) is found.

To obtain energy-dependent flux upper limits, we divide the whole energy range of $500\,{\rm MeV} - 500\,{\rm GeV}$ into 20 logarithmically-spaced energy bins. For each energy bin, the background sources are fixed and a power-law spectral model ($dN/dE \propto E^{-\Gamma}$) with $\Gamma$=2 \cite{fermi15dsph} is used to model the putative dSph source. 
We get the upper limits on flux at a $95\%$ confidence level when the log-likelihood ($\ln \mathcal{L}$) changes by 1.35 \cite{ROLKE2005493}.
The energy-dependent flux upper limits for the 4 sources with the biggest J-factors are shown in Fig.~\ref{fig1}, while the flux upper limits for other dSphs in our sample are presented in Appendix~\ref{app2}.

%*******************table.1****************************

\begin{table}
    \caption{The information of the considered dSphs.}
    \tabcolsep=0.4cm
    \begin{tabular}{lcccccc}
        \hline
        \hline
        Name & $(l,b)$ & $\log_{10}{(J)}^1$ & $\log_{10}(J_{\rm flat})^2$ &TS\\
        & [deg] & [$\log_{10}\rm (GeV^{2}cm^{-5})$] &[$\log_{10}\rm (GeV^{2}cm^{-5})$]& \\
        \hline
        Bootes I&(358.1, 69.6)         &$18.8\pm0.22$ &16.54 / 16.84&0.00\\
        CanesVenatici II&(113.6, 82.7) &$17.9\pm0.25$ &17.48 / 17.94&2.51\\
        Carina & (260.1, -22.2)        &$ 18.1\pm0.23$&17.90 / 18.15&0.00\\
        ComaBerenices &(241.9, 83.6)   & $19.0\pm0.25$&18.56 / 18.85&0.00\\
        Draco &(86.4, 34.7)            & $18.8\pm0.16$&18.87 / 19.00&0.00\\
        Fornax &(237.1, -65.7)         & $18.2\pm0.21$&18.07 / 18.29&1.73\\
        Hercules &(28.7, 36.9)         & $18.1\pm0.25$&16.56 / 17.28&4.53\\
        Leo II &(220.2, 67.2)          & $17.6\pm0.18$&17.41 / 17.49&0.00\\
        Leo IV &(265.4, 56.5)          & $17.9\pm0.28$&16.48 / 16.90&0.00\\
        Sculptor &(287.5, -83.2)       & $18.6\pm0.18$&18.56 / 18.80&0.00\\
        Segue I &(220.5, 50.4)         & $19.5\pm0.29$&19.26 / 19.66&2.74\\
        Sextans &(243.5, 42.3)         & $18.4\pm0.27$&17.77 / 18.04&0.00\\
        UrsaMajor II &(152.5, 37.4)    & $19.3\pm0.28$&19.15 / 19.77&0.00\\
        UrsaMinor &(105.0, 44.8)       & $18.8\pm0.19$&18.96 / 19.47&1.20\\
        Willman I &(158.6, 56.8)       & $19.1\pm0.31$&19.14 / 19.54&8.16\\
        CanesVenatici I &(74.3, 79.8)  & $17.7\pm0.26$&17.15 / 17.46&0.00\\
        Leo I& (226.0, 49.1)           & $17.7\pm0.18$&17.75 / 17.89&1.53\\
        UrsaMajor I&(159.4, 54.4)      & $18.3\pm0.24$&18.11 / 19.11&0.00\\
        Reticulum II&(266.3, -49.7)    & $19.4\pm0.40$&18.50 / 19.06&7.36\\
        \hline
    \end{tabular}
    \begin{tablenotes}
    \item { $^1$ J-factors derived from stellar kinematics. For Reticulum II, it is taken from \cite{Simon:2015fdw}, others are from \cite{fermi15dsph}}.
    \item { $^2$ {J-factors taken from \cite{Sanders:2016eie}, which have accounted for the flattening of the dSphs. The left and right values are for the oblate and prolate cases, respectively.}}
    \end{tablenotes}
    \label{tab1}
\end{table}

%################################################################################

To improve the sensitivity, we also perform a stacking analysis of our sample. 
The stacking analysis can give a better sensitivity by merging the observations of multiple sources.
We sum the data (i.e. the count cubes in the binned likelihood analysis) of these sources together.
The likelihood is evaluated by
	\begin{equation}
		\ln\mathcal{L}= \sum_i N_i\ln M_i-M_i-\ln N_i!
		\label{like}
	\end{equation}
where $M_i$ and $N_i$ is the model-predicted and observed counts in each pixel, respectively,
and the index $i$ runs over all energy and spatial bins.
For our stacking analysis, $M_i=\sum_k m_{i,k}$ and $N_i=\sum_k n_{i,k}$ with index $k$ summing over all target sources.
The $m_i$ is obtained using the {\tt gtmodel} command in the {\tt Fermitool} software, while the $n_i$ is the counts cube obtained using the {\tt gtbin} command.
The model map $m_i$ is related to the DM model parameters and the dSph J-factor, $m_i=m_i({\Phi}(\left<\sigma v\right>,m_\chi,J))$. The DM annihilation flux ${\Phi}(\left<\sigma v\right>,m_\chi,J)$ is implemented in the analysis with a {\tt FileFunction} spectrum\footnote{\url{https://fermi.gsfc.nasa.gov/ssc/data/analysis/scitools/source_models.html\#FileFunction}} (see Sec.~\ref{sec:constr} for the calculation of the expected DM spectrum).
Note that the method used here is not the same as the commonly-used combined likelihood analysis \cite{fermi11dsph,tsai13dsph,fermi14dsph,fermi15dsph}. 
The advantage of our method is that it can effectively avoid the loss of sensitivity due to the uncertainty of J-factor.
For comparison, we also perform the combined analysis (see Appendix~\ref{app1} for details) which gives similar results and the same conclusion.

%{\bf counts map, model map, residual map}

We use the above method to scan a series of DM masses.
Since we are focusing on the i2HDM model that can commonly interpret the W-boson mass anomaly, the GC GeV excess and the antiproton excess, we scan the masses from 45 GeV to 140 GeV for DM annihilation channel $SS\rightarrow WW^*$.
%The TS values as a function of DM mass are presented in Fig.~\ref{fig:1}. 
For all  dSphs, no significant (i.e. $\rm TS>25$) signals are found in our analyses.  We show our results in Table~\ref{tab1}.
One reason why our analysis does not give a relatively high TS value (i.e. ${\rm TS}>10$) as in previous results \cite{gs15ret2,hooper15ret2,Drlica-Wagner:2015xua,fermi2017dsph,2018PhRvD..97l2001L,hoof20} for the source Reticulum II is that we have added an additional source 0.15$^\circ$ away from Reticulum II into the background model. In our previous work we have shown that the excess from the direction of Reticulum II offsets this dSph by $\sim0.15^\circ$ and may be not associated with it \cite{ls21}. If not subtracting this excess, we will obtain weaker constraints on the model, but the conclusions would not change.
In the stacking analysis, the TS value for the total emission from the dSphs is $\sim0.2$.

\section{Constraints on the cross section of $SS\rightarrow WW^*$ annihilation}
\label{sec:constr}

\begin{figure}[h]
	\includegraphics[width=0.6\textwidth]{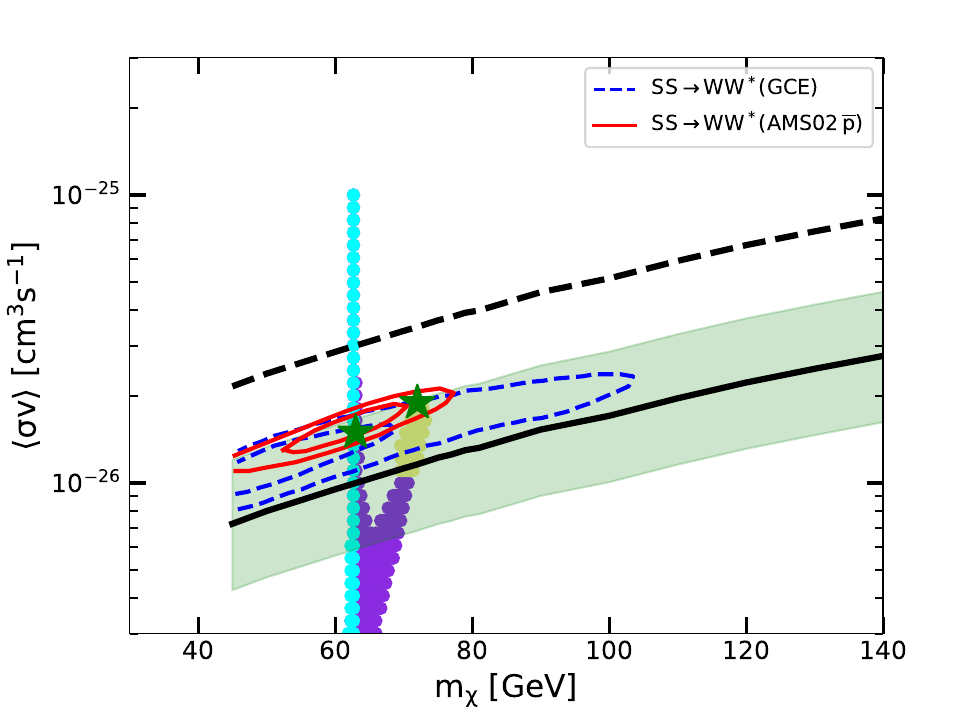}
    \caption{Upper limits at a 95\% confidence level on the cross section of the $SS\rightarrow WW^*$ annihilation (thick black line). The contours demonstrate the favored DM parameters of $SS\rightarrow WW^*$ via fitting to the antiproton and GC excess data ($1\sigma$ and $2\sigma$ from inside to outside, extracted from \cite{zcr22}). The colored regions represent the allowed parameters to interpret the W-boson mass of CDF II derived in \cite{fyz22}. Three colors are for different DM production mechanisms in the early Universe: SA coannihilation (purple), Higgs resonance (cyan), and $SS\rightarrow WW^*$ annihilation (yellow). The parameters that can simultaneously interpret the W-boson mass anomaly in the i2HDM model are around the star symbols. {The shaded band illustrates the variation of the limits considering the flattening of the dSphs \cite{Sanders:2016eie}.} Since our constraints are relatively marginal, if the constraints are weakened by (e.g.) a factor of 3 (dashed line) due to the uncertainties of the J-factors, the parameters of interest can not be ruled out (see the discussion in Sec.~\ref{sec:constr} for details).}
	\label{fig2}
\end{figure}

Since we do not find any signals from the directions of these dSphs (both individual and stacking analyses), we test whether the i2HDM parameters accounting for the three anomaly/excesses are in tension with the dSph's null results.
We derive the upper limits on the cross section $\left<\sigma v\right>_{SS\rightarrow WW^*}$ for a series of DM masses of 45-140 GeV.
The method described in the Sec.~\ref{sec:analysis} are used to get the constraints on the cross section.  
The expected $\gamma$-ray flux from DM annihilation reads
\begin{equation}
	{\Phi}(E_{\gamma})={\frac{\left<{\sigma}v\right>}{8{\pi}m_{S}^{2}}\frac{dN_{\gamma}}{dE_{\gamma}}\times J},
\end{equation}
where ${m_{S}}$ and ${\left<{\sigma}v\right>}$ are the DM particle mass and the velocity-averaged DM annihilation cross
section.
The term 
\begin{equation}
	J={\int_{\rm ROI}}\int_{\rm l.o.s}{\rho}^{2}(r)dld{\Omega}
\end{equation}
is the so-called J-factor, which can be determined through stellar kinematics.
The J-factors of our sources are listed in Table~\ref{tab1} which are extracted from Ref.~\cite{fermi15dsph,Simon:2015fdw}.
The $dN_{\gamma}/dE_{\gamma}$ is the differential $\gamma$-ray yield per annihilation. 
For the $SS\rightarrow WW^*$ channel and when the DM mass is $m_{\rm S} < m_{\rm W}$, the off-shell annihilation into $WW^*$ is considered and 
we use the spectra the same as those in \cite{zcr22}, which is simulated with {\tt MadGraph5\_aMC@NLO} \cite{MadGraph5} and {\tt PYTHIA8} \cite{pythia8}. 
For the DM mass $m_{\rm S} > m_{\rm W}$, the DM spectra are obtained from PPP4DMID \cite{Cirelli:2010xx}.

The obtained constraints are shown in Fig.~\ref{fig2}. The i2HDM parameters that can simultaneously interpret the W-boson mass anomaly, the GC GeV excess and the GeV antiproton excess are the overlapping parameters of the red and blue contours and the colored points, and are within very small regions around the star symbols. We find that our upper limits (the thick black line) are below the favored  parameters, suggesting that the dSph observations seem to be able to exclude such a model. 

We would like to give some comments and discussions on this result. Firstly, it should be noted that our results exclude only the parameter space that can {\it simultaneously} explain the three anomaly/excesses, but not the whole i2HDM model of accounting for the W-boson mass anomaly. For the i2HDM model with cross section $\left<\sigma v\right> \lesssim 10^{-26}\,{\rm cm^3s^{-1}}$, it can still be used to explain the CDF II W-boson mass. In fact, as is shown in Fig.~\ref{fig2}, our limits mainly exclude the parameters related to the GC excess (blue contour) and the GeV antiproton excess (red contour). Such a conclusion is supported by some previous works, which also showed that the best-fit GC excess parameters are not favored by the dSph observations \cite{fermi15dsph,fermi2017dsph,hoof20,Hess:2021cdp}. 

However, the dSph constraints (including our results) are reliant on the accuracy of the J-factor measurements, but actually there are substantial uncertainties in current J-factor measurements. 
As seen in \cite{liang2016dl}, for the same data analysis, the use of J-factors provided by different groups can lead to results that differ by a factor of several.
Moreover, some studies have revealed that the existing estimations of J-factors are not conservative. 
For example, Ref.~\cite{Ichikawa:2016nbi} figures out that considering the contamination of foreground stars may decrease the J-factors by a factor of 3 (thus weakening the dSph constraints by a factor of 3); numerical simulations of galaxy formation that take into account the baryonic effects also point out that the flux of DM annihilation from dSphs may be much lower than that of the Galactic center \cite{2021MNRAS.501.3558G}, thus making the null detection in our analysis reasonable even if the GC GeV excess was a true DM signal.
{Another source of uncertainty that should also be assessed is the flattening of the dSphs. Our benchmark limits (the thick black line in Fig.~\ref{fig2}) is derived based on the J-factors that are calculated assuming spherical models of the dSphs.
It is argued that the flattening of the galaxies will lead to J-factor values different by tens of percent \cite{Sanders:2016eie} (the J-factors that consider the flattening are also listed in Table I).
The shaded band in Fig.~\ref{fig2} illustrates the variation of the limits on $\left<\sigma v\right>$ due to the flattening, where the upper (bottom) bound assumes that the dSphs are all oblate (prolate).}
Finally we also do not consider the extension of dSphs in our data analysis, instead treating dSphs as point-like sources. Ref.~\cite{DiMauro:2022hue} has shown that taking into account the extension of dSphs, the limits on DM parameters will be weakened by a factor of $\sim1.5-2.5$.
Considering all these factors, the constraints from dSphs are possible to be several times weaker and thus unable to constrain the model (see the dashed line in Fig.~\ref{fig2}).

Therefore, it will be crucial to use better observations of dSphs from future observatories (e.g., Vera C. Rubin Observatory \cite{2009arXiv0912.0201L}) to accurately determine the J-factor values.
The next generation gamma-ray telescopes with a sensitivity several times better (e.g., VLAST \cite{2022AcASn..63...27F}) is also promising to give a reliable answer to the problem.

\section{Summary} 
The new measurement of W-boson mass by the CDF collaboration shows that the mass deviates from the Standard Model prediction with a significance of $>7\sigma$ \cite{wboson}.
This result indicates there may exist new physics beyond the SM (see however the recent W-boson mass measurements by the ATLAS collaboration \cite{ATLAS-CONF-2023-004,ATLAS:2017rzl}, which do not support the results by CDF). 
Ref. \cite{fyz22,zcr22} proposed that the inert two Higgs doublet model (i2HDM) can well explain the new W-boson mass. More encouragingly, they found that this model can also explain both the GeV gamma-ray excess in the Galactic center and the GeV antiproton excess with common parameters. The gamma rays and cosmic rays are produced through a $SS\rightarrow WW^*$ annihilation with $S$ the lightest stable particle in the i2HDM which can play the role of DM. 

In this paper, we have tested the possible common origin of the three anomaly/excesses by analyzing the {\it Fermi}-LAT observations of dSphs using the dSph J-factors reported in Refs.~\cite{fermi15dsph,Simon:2015fdw}.
The Milky Way dSphs are an ideal source population to test DM models due to their large J-factors and low gamma-ray background. 
We do not find any significant signal in both the single-source and the stacking analyses. Based on the null results, we place constraints on the cross section of the self-annihilation of the DM paticle $S$.
We find that our constraints seems to be able to exclude, at a 95\% confidence level, the favored parameters reported in Ref.~\cite{zcr22} that can simultaneously interpret the W-boson mass anomaly, the GC excess and the antiproton excess. 
However, we point out that there exist uncertainties in our exclusion limits and we still cannot reliably claim that the common origin has been excluded.
The reason is that the model parameters are only marginally excluded while the exclusion line we obtained relies on the accuracy of the J-factor measurements which however suffers from considerable uncertainties. 
It is expected that future detectors with higher sensitivity (e.g., VLAST) will be able to solve this problem (either reliably exclude the model or detect a signal). 

\begin{acknowledgments}
We thank Yizhong Fan, Ziqing Xia, Zhaohuan Yu and Qiang Yuan for providing us the DM annihilation spectra of $WW^*$ channel and for the helpful communications. This work is supported by the National Key Research and Development Program of China (No. 2022YFF0503304) and the Guangxi Science Foundation (grant No. 2019AC20334).
\end{acknowledgments}

\bibliographystyle{apsrev4-1-lyf}
\bibliography{refs.bib}

\appendix
\section[\appendixname~\thesection]{Combined likelihood analysis}
\label{app1}
\begin{figure}[h]
		\includegraphics[width=0.5\textwidth]{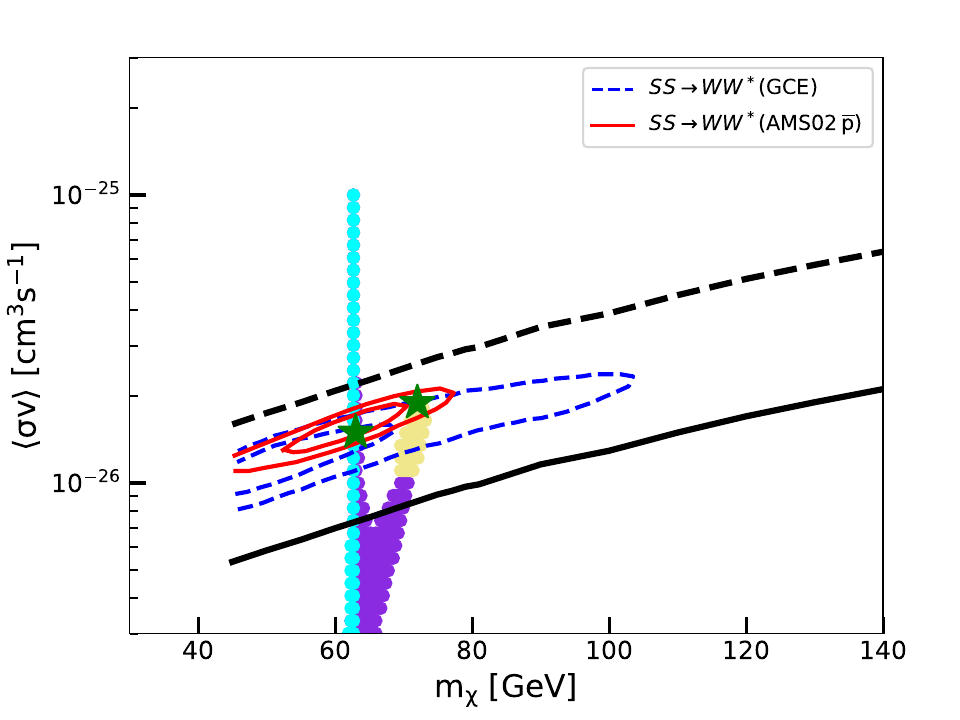}
        \caption{{Upper limits based on a combined likelihood analysis which gives results consistent with those in Fig.~\ref{fig2}}.}
		\label{figA1}
	\end{figure}

In order to verify the stacking analysis employed in our main text (the data and model are respectively summed over all sources before calculating the likelihood), here we also use the combined likelihood analysis (the likelihoods are calculated source by source and then summed together to obtain a total likelihood), which has been widely used in previous studies \cite{fermi11dsph,tsai13dsph,fermi14dsph,fermi15dsph}, to derive the corresponding results for comparison.
We divide the data in the energy range of 500 MeV-500 GeV into 20 logarithmically-spaced energy bins. For each energy bin $k$ we vary the scale parameter of the dSph component and derive the relation between the likelihood $\mathcal{L}_k$ and the target flux within the $k$th bin $f_k$, $\mathcal{L}_k(f_k)$ (namely the likelihood profile).
The total likelihood considering all energy bins for the DM model with parameters $\left<\sigma v\right>$ and $m_S$ is given by:
\begin{equation}
    \mathcal{L}' (\left<\sigma v\right>,m_S,J) = \prod_k{\mathcal{L}}_{k}(f_k(\left<\sigma v\right>, m_S,J)).
\label{eq:a1}
\end{equation}
To combine all sources in the sample, the combined likelihood is
\begin{equation}
    \tilde{\mathcal{L}}(\left<\sigma v\right>,m_S) = \prod_i{\mathcal{L}'_i}(\left<\sigma v\right>, m_S,J_i)\times\mathcal{L}_{\rm J} (J_i\,|\,J_{{\rm obs},i},\sigma_i)
\end{equation}
with $\mathcal{L}'_i$ the Eq.~(\ref{eq:a1}) and for the $i$th source. In the combined analysis, we also consider the statistical uncertainty in the J-factors, which is incorporated into the likelihood through the term
\begin{equation}
	\mathcal{L}_{\rm J} (J\,|\,J_{{\rm obs},i},\sigma_i) = \frac{1}{\ln(10)J_{{\rm obs},i}\sqrt{2\pi}\sigma_i}{\rm e}^{-\left[\log_{10}(J)-\log_{10}(J_{{\rm obs},i})\right]}
\end{equation}
 where $J_{{\rm obs},i}$ and $\sigma_i$ are the measured J-factor and its uncertainty for the source $i$.  For a given $m_S$, the upper limit on $\left<\sigma v\right>$ at a 95\% confidence level is derived by requiring $\ln \tilde{\mathcal{L}}$ to change by 1.35. We show the constraints based on the combined analysis in Fig.~\ref{figA1}. 

\newpage
\section[\appendixname~\thesection]{Bin-by-bin flux upper limits for other sources}
\label{app2}
	\begin{figure}[h]
	\includegraphics[width=0.98\textwidth]{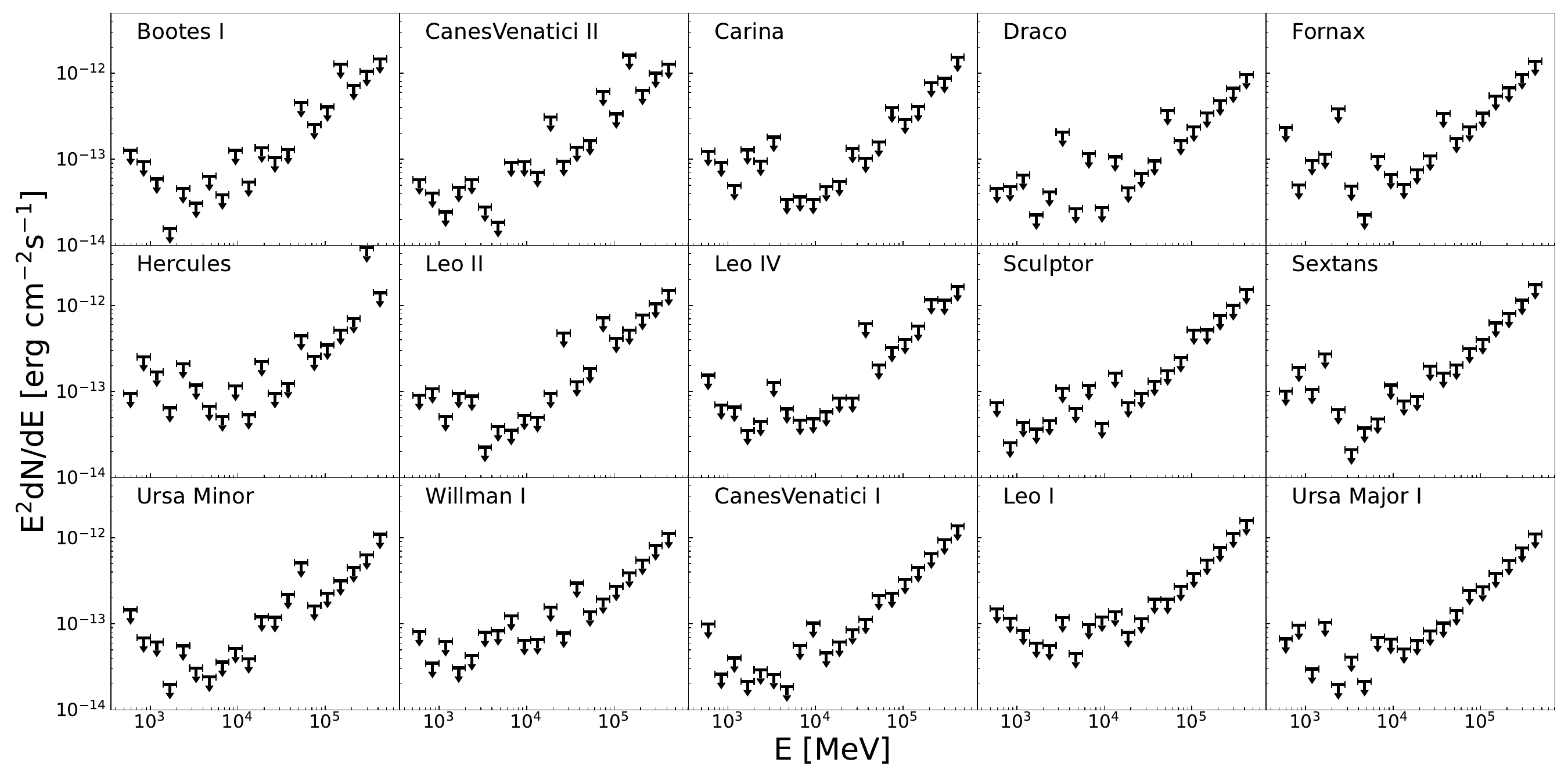}
		\caption{Bin-by-bin flux upper limits at a 95\% confidence level for the rest dSphs considered in this work, as a supplement to Fig.~\ref{fig1}.}
		\label{figA2}
	\end{figure}

\end{document}